\documentclass[fleqn,twoside]{article}
\usepackage{espcrc2}

\usepackage{graphicx}
\usepackage[figuresright]{rotating}

\newcommand{\AmS}{{\protect\the\textfont2
  A\kern-.1667em\lower.5ex\hbox{M}\kern-.125emS}}

\title{European Perspectives for Electron-Nucleon Scattering \\
       at the Luminosity Frontier}

\author{Wolf-Dieter Nowak\address[MCSD]{DESY Zeuthen, Platanenallee 6, D-15738
        Zeuthen, Germany}
        \thanks{Invited talk at the Workshop `The Spin Structure of the Proton
         and Polarized Collider Physics', Trento/Italy, July 23-28, 2001}
}
\begin{document}

\begin{abstract}
European perspectives are discussed on fixed-target electron or positron
scattering experiments using polarized and unpolarized beams and targets 
in various combinations. The goal envisioned 
is a deep and complete understanding of the momentum and spin structure 
of hadrons in the context of Quantum Chromo Dynamics, based on measurements
at moderate and low photon virtualities. This program can be realized by
performing electron-nucleon experiments with high resolution and
high luminosity, at beam energies of at least 25-30 GeV.

\vspace{1pc}
\end{abstract}

\maketitle

\section{INTRODUCTION}

Over more than two decades the momentum and spin structure of single
partons in the nucleon has been investigated by now, preferentially 
using charged leptons as probe(s). A great variety of measurements, 
performed in fixed-target and collider experiments, 
turned Quantum Chromo Dynamics (QCD) from a candidate theory into 
the widely accepted field theory of strong interactions. Large enough
photon virtualities $Q^2$ were required to successfully test the 
abilities of perturbative QCD to describe short-range phenomena. 
In contrast, the description of long-range phenomena and especially 
of parton correlations in hadrons, over a broad range in $Q^2$, is 
still in its infancy. It may be expected that contemporary theoretical 
developments will at some moment evolve into a calculable field 
theoretical description of hadronic structures.
It is hence vital to maintain in this process 
a permanent confrontation of theoretical ideas to measurements. \\

In this paper possible ways are discussed how to acquire in the future
adequately precise experimental data to test different approaches and 
to eventually be able to select the correct method
describing hadronic structure on the basis of fundamental interactions.

\section{PHYSICS MOTIVATION}

Deep inelastic lepton-nucleon scattering (DIS) continues to be a 
very successful tool to explore perturbative QCD. On the one hand, a
wealth of inclusive data from `unpolarized' fixed-target and collider 
experiments at CERN, DESY and FNAL allowed a 
rather accurate determination of the quark distributions 
$q(x,Q^2)$ through precise QCD fits of the unpolarized nucleon structure 
function $F_2(x, Q^2)$. Here $x$ is the Bjorken scaling variable 
describing the constituent's momentum fraction in a fast moving hadron. 
On the other hand, results from several `polarized' fixed-target 
experiments at CERN, DESY and SLAC measuring double spin asymmetries 
in inclusive DIS cross sections led to 
the determination of the polarized structure function $g_1(x, Q^2)$ 
with modest accuracy. Although this data is limited to moderate values 
in $Q^2$, it constrains the contribution of the quark spin 
to the nucleon's helicity on the 10-20\% level of accuracy.

Recent and near-future measurements at CERN (COMPASS), DESY (HERMES) and
JLab are changing the emphasis from inclusive DIS to semi-inclusive 
(SIDIS) measurements. The main thrust of COMPASS will be a determination 
of the polarized gluon distribution $\Delta G(x)$ \cite{DG_G_compass}, 
over a limited range of $x$ with about 10\% accuracy 
(cf. Fig.~\ref{fig:DeltaG_G}). Helicity distributions $\Delta q(x)$ will be 
available from HERMES, on the few \% level for valence quarks and for 
the first time for sea quarks \cite{MarcDIS2001}. A first measurement
of the transversity $\delta q(x)$ \cite{RalstonSoper,ArtruMekhfi,JaffeJi}, 
characterizing the 
distribution of the quark's spin in a transversely polarized nucleon, 
is the main goal of HERMES in 2002-03 \cite{KNO}. Accomplishing a 
complete understanding of the nucleon structure on the twist-2 level 
requires precise experimental data for all parton distribution 
functions mentioned above.

A few years ago Generalized Parton Distributions (GPDs) 
\cite{Mueller,Radyushkin,Ji,Bluemlein}, a set of non-perturbative 
distribution functions, were introduced in the description of several 
processes ranging 
from inclusive to hard exclusive scattering. GPDs embody ordinary 
parton distribution functions and nucleon form factors. They depend 
on $Q^2$, two longitudinal momentum fractions and $t$, the momentum 
transfer at the nucleon vertex. GPDs can in principle be 
revealed from a set of different cross sections and asymmetries for
various exclusive processes.

The recent strong interest in GPDs was stimulated
by the finding \cite{Ji} that the total angular momentum carried by
partons of a certain type is given by the second moment of the sum of the
unpolarized GPDs describing this type of partons, in the limit of small $t$. 
The total angular momenta carried by quarks and gluons in the nucleon, 
$J_q$ and $J_g$, constitute the hitherto missing pieces in the puzzle 
representing the momentum and spin structure of the nucleon. 

Nowadays hard exclusive lepton-nucleon processes appear as a viable 
tool to access quark GPDs by measuring a photon or a scalar, pseudo-scalar 
or vector meson in the final state. Very recently, first experimental
results on Deeply Virtual Compton Scattering (DVCS) have been obtained.
The fixed-target data from HERMES \cite{HermesDVCS} and CLAS (JLab) 
\cite{ClasDVCS} exhibit significant asymmetries in the distribution 
in the azimuthal angle between lepton and hadron planes, thereby 
confirming the expected interference between the dominating DVCS 
handbag diagram and the background Bethe-Heitler process. H1 measured
the DVCS cross section at the HERA collider \cite{H1DVCS} and found
agreement with QCD-based calculations \cite{FFS-DD}.

Even after completion of the present series of lepton-nucleon SIDIS 
measurements\footnote{The RHIC $pp$ spin program \cite{RHIC} will provide 
precise and 
complementary data on $\Delta G$ and $\delta q$. The comparison of these
measurements to SIDIS data will allow for very important consistency checks,
in particular for a test of factorization in spin-dependent processes.}
both helicity and transversity quark distributions as well as the 
polarized gluon distribution will remain clearly inferior in accuracy 
compared to the unpolarized distributions. 
Concerning GPDs it is generally 
agreed that because of their complexity every attempt to really measure 
them will require a huge common effort of theory and experiment. \\

\section{EXPERIMENTAL FACILITIES}

The above envisioned `complete' determination of the momentum and spin
structure of the nucleon (and other hadrons) clearly calls for new
experimental facilities which, in comparison to existing experiments,
include
\begin{itemize}
\item an accelerator with a high duty factor of at least 5-10\% allowing
      for hitherto unprecedented luminosities of $10^{35}$ cm$^{-2}$s$^{-1}$ 
      or more;
\item beams of electrons and, if possible, positrons\footnote{In the 
      following only the term `electron' will be used, although most 
      experiments can either be done with electrons or positrons, or
      with muons in a future high intensity muon collider or neutrino
      factory.} 
      with an energy definition better than $\cal{O}$~$(10^{-4})$ and 
      a high degree of polarization;
\item beam energies above 25-30 GeV to reach the perturbative 
      regime of a few GeV$^2$ in $Q^2$, over the full kinematically  
      accessible $x$-range;
\item radiation-tolerant (cryogenic) solid-state targets that can 
      achieve high degrees of longitudinal and transverse polarization;
\item a spectrometer that combines very fast, $\cal{O}$(1-2 ns) 
      triggering with high-precision tracking and full particle 
      identification.
\end{itemize}
These requirements are the basis for two proposals, TESLA-N and ELFE@DESY,
described in the addendum to the Technical Design Report \cite{TESLA-TDR} 
for the TESLA project, a 250 $\times$ 250 GeV 
super-conducting linear $e^+e^-$ collider. 

In the case of TESLA-N \cite{TESLA-N} it is foreseen to inject separately 
produced (polarized) electron bunches into the positron arm of the TESLA 
machine and to extract them at one of several possible locations,
thereby realizing a beam energy range between 30 and 250 GeV. 
The (polarized) electrons would be directed towards a separate 
experimental hall. Here a new experiment would combine a (polarized)
solid-state target with a new multi-stage spectrometer required due to
the large energy range of the produced particles.

\begin{figure}[htb]
\vspace{-0.6cm}
\centering
\includegraphics[width=7.2cm]{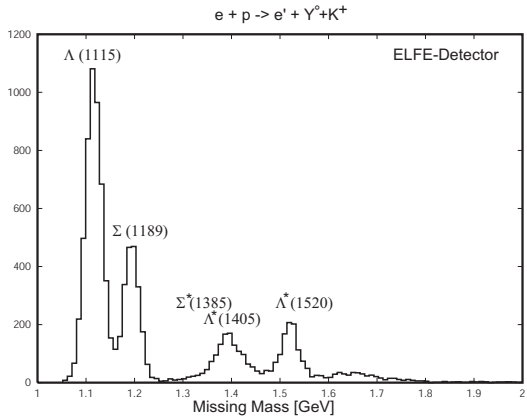}
\vspace{-0.7cm}
\caption{\it Illustration of the missing mass resolution of a ELFE@DESY-type 
             spectrometer \cite{TESLA-ELFE,DvHpriv}.}
\label{fig:resolELFE}
\end{figure}

In the case of ELFE@DESY \cite{TESLA-ELFE}, separate electron bunches would be 
accelerated in the first part of TESLA to about 30 GeV, transported 
back and subsequently injected
into the HERA-$e$ ring. The latter would serve as stretcher to 
increase the low TESLA duty cycle of 0.5\% by a substantial factor, 
whereby keeping a precise energy definition of the beam. In the
experiment a (polarized) solid-state target would be combined with a
new high-resolution spectrometer that will allow for a clear separation 
of the nucleon ground state from isobars. Its projected performance is
illustrated for the exclusive channel $e p \rightarrow e K^+ Y^0$
in Fig.~\ref{fig:resolELFE} \cite{DvHpriv}, assuming that electron and 
kaon were measured in the spectrometer while the missing mass of the 
hyperon had to be calculated. 
\vspace{-0.6cm}

Both concepts may be profitably combined: The technical solution 
suggested for the beam part of TESLA-N in conjunction with a (polarized) 
solid-state target and a high-resolution ELFE@DESY-type spectrometer would 
be ideal to pursue the complex physics programme outlined above. This 
option could also be realized sharing a lower-energy, 30-50 GeV 
TESLA-like accelerator that is presently discussed to feed the X-Ray 
Free Electron Laser facility \cite{TESLA-FEL}.

In the case that HERA running should be extended beyond 2006, a 
certain part of this physics programme could already be realized 
with an upgraded HERMES apparatus.

\begin{figure}[ht]
\vspace{-1.3cm}
\includegraphics[width=8cm]{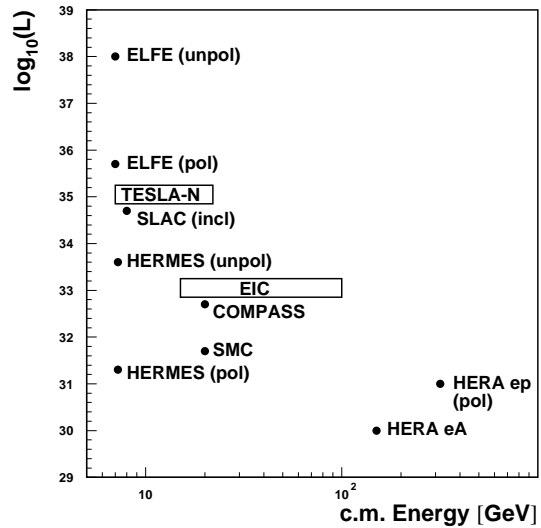}
\vspace{-1.5cm}
\caption{\it Luminosity versus c.m. energy for existing and (possible) 
         future facilities. 
}
\label{fig:lumi_E}
\end{figure}

The envisioned luminosities of both projects, TESLA-N and ELFE@DESY, are 
depicted in Fig.~\ref{fig:lumi_E} compared to existing and other possible
future facilities. The planned TESLA duty cycle of 0.5\%
is the only reason for the lower luminosity limit of TESLA-N in 
comparison to ELFE@DESY. In any case a luminosity limit 
of a few times $10^{35}$ cm$^{-2}$s$^{-1}$ is imposed by the target 
cryogenic system when using a polarized solid-state target. This
determines also the maximum possible luminosity for a polarized 
combined TESLA-N/ELFE@DESY experiment once a duty cycle 
of a few \% is available.
\vspace{-0.7cm}

Note that the effective luminosity of any experiment using a polarized
solid-state 
target, such as the polarized NH$_3$-target of SMC, is lower by a factor 
of about 25 compared to an electron-proton collider (HERA, EIC) or an 
internal-proton-target experiment (HERMES), due to the large dilution 
factors involved.
This dilution effect is not included in Fig.~\ref{fig:lumi_E}.

\section{PROJECTIONS FOR FUTURE \\ MEASUREMENTS}

\subsection{Polarized Gluon Distribution}

\begin{figure}[htb]
\vspace{-1cm}
\centering
\includegraphics[width=8cm]{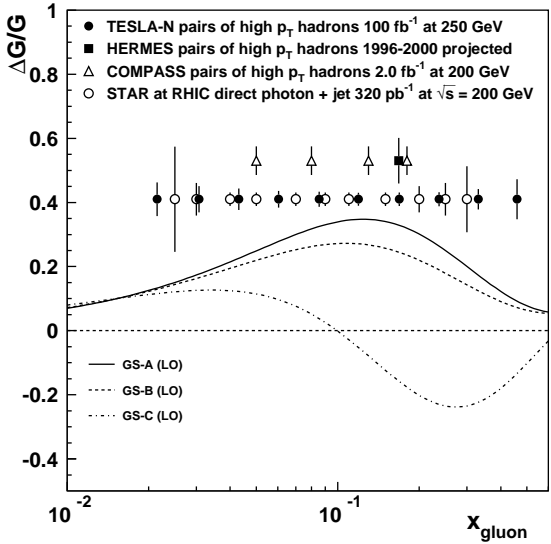}
\vspace{-1.5cm}
\caption{\it Projected statistical 
          accuracies for the measurement of $\Delta G(x)/G(x)$ at 
          TESLA-N, based on an integrated luminosity of 100 fb$^{-1}$, 
          in comparison to projections from COMPASS~\cite{DG_G_compass},
          HERMES~\cite{DG_G_hermes} and RHIC~\cite{RHIC}.
          The phenomenological predictions~\cite{GSLO} were calculated 
          for Q$^2$ = 10 GeV$^2$ (taken from Ref.~\cite{TESLA-N}).}
\label{fig:DeltaG_G}
\end{figure}

\vspace{1ex}
In Fig.~\ref{fig:DeltaG_G} representative projections are shown for 
future results on the polarized gluon distribution.
Highly precise measurements as planned for 
\vspace{-0.8cm}
RHIC and proposed for 
TESLA-N are eventually needed to map out the full $x$-dependence 
of $\Delta G(x)$ with reasonable precision. The soon expected
COMPASS results (including the not shown results from open charm
production) will provide very valuable information on $\Delta G(x)$ due 
to their completely different systematics as compared to RHIC. 
No projections exist up to now for the accuracy of measurements of 
$\Delta G(x, Q^2)$.

The RHIC projections shown in Fig.~\ref{fig:DeltaG_G} were calculated 
for a given final state
and $\sqrt{s}=200$ GeV, i.e. for $Q^2 = \cal{O}$(100 GeV$^2$). They
will exhibit larger error bars, when evolved down to $Q^2=10$ GeV$^2$.
At the same time, $\Delta G(x)$ will be determined from other final states 
and at two different energies. Still, a measurement of 
$\Delta G(x)$ at RHIC is not without problems in view of the large 
theoretical uncertainties in direct photon and heavy quark-pair 
hadroproduction. 
Hence a statistically highly precise and systematically independent 
determination of $\Delta G(x, Q^2)$ through an electron-nucleon scattering 
experiment is clearly needed. 

\subsection{Transversity Distributions}

\begin{figure}[htb]
\centering
\vspace{-1.2cm}
\includegraphics[width=7.6cm]{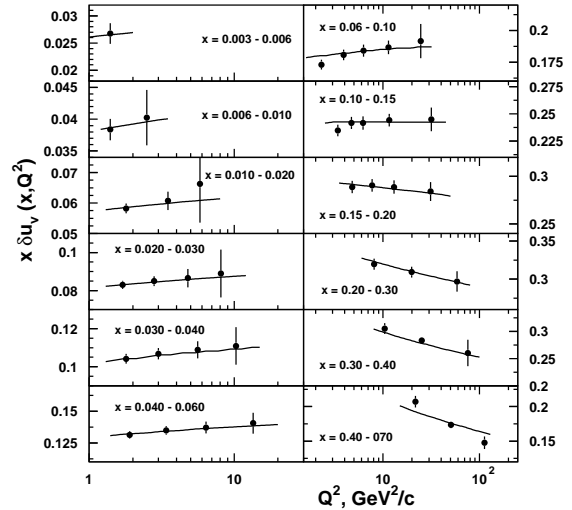}
\vspace{-1.4cm}
\caption{\it The valence $u$-quark 
             transversity distribution as a function of $x$ and $Q^2$ 
             as it would be measured at TESLA-N, based on an integrated 
             luminosity of 100 fb$^{-1}$. The curves show the LO 
             $Q^2$-evolution of the $u_v$-quark transversity distribution 
             through a fit to the simulated asymmetries (taken from 
             Ref.~\cite{TESLA-N}).}
\label{uvalTransv-5mrad}
\end{figure}

\vspace{1ex}
In Fig.~\ref{uvalTransv-5mrad} the projected statistical accuracy is 
shown for a measurement of the $(x, Q^2)$-dependence of the $u_v$-quark 
transversity distribution through single-spin azimuthal cross-section 
\vspace{-0.7cm}
asymmetries in semi-inclusive charged pion production at TESLA-N 
\cite{TESLA-N}. In the simulated 
analysis it was assumed that transversity and helicity distributions 
coincide at photon virtualities of the order of 1 GeV$^2$ and 
valence-like $x$-values. Fig.~\ref{uvalTransv-5mrad} shows that
a broad $x$-range ($0.003 < x < 0.7$) 
can be accessed while $Q^2$ is in between 1 and 100 GeV$^2$,
with an impressive statistical accuracy over almost the full range.
Because of $u$-quark dominance in pion electro-production a lower
accuracy is attained in the reconstruction of the other transversity 
distributions, $\delta d_v$, $\delta \bar{u}$, and $\delta \bar{d}$.

Measuring the transversity distributions for quarks and anti-quarks
of a given flavor would allow for the first time to determine the 
tensor charge $\delta \Sigma(Q^2)$ of this flavor. This would give
access to the hitherto unmeasured chirally-odd operators in QCD which
are of great importance to understand the role of chiral symmetry in the
structure of the nucleon \cite{Jaffe97}.
A fit to a simulated data sample, as described in Ref.~\cite{TESLA-N},
also provides projections for the accuracy of the $u$- and $d$-quark
tensor charges: $0.88 \pm 0.01$ and $-0.32 \pm 0.02$ at a scale of 
1~GeV$^2$. 
Note that the absolute values of the tensor charges are 
determined to a large extent by the input distributions used, although 
their values are rather close to those predicted by lattice 
QCD calculations.

\subsection{Generalized Parton Distributions}

\vspace{1ex}
The following discussion focuses on DVCS as a tool to derive
$J_q$ through the determination of GPDs. HERMES plans for 2004-2006 
to use an unpolarized target in conjunction with a new 
recoil detector \cite{HERMESrecoil}. 
As can be seen from the anticipated statistical precision, evaluated
in Ref.~\cite{KoNoDVCS}, rather accurate data can be expected
on the azimuthal dependence of the asymmetries in 
beam helicity and beam charge. In addition, for the first 
time the dependence of these asymmetries on $t$, the third variable on 
which GPDs depend, may become accessible. Even a first approximate
determination of the GPD $H^u$ may be tried on the basis of $u$-quark
dominance.

Explicit formulae for azimuthal DVCS asymmetries on the twist-2 level
are given e.g. in Ref.~\cite{BeMueNieSch}. In case of measurements
with a polarized beam and an unpolarized target the generalized parton 
distribution function $H$ dominates, while the two other involved 
functions, $\tilde{H}$ and $E$, are suppressed relative to $H$ by 
kinematic factors.

Determining through Ji's relation \cite{Ji} the total angular momentum 
$J_q$ carried by quarks of flavor $q$ requires data
not only on the helicity-conserving GPD $H$, but as well on the 
helicity-flip GPD $E$. Access to $E$ may be achieved through DVCS
measurements with an unpolarized beam and a transversely 
polarized target. Also, important additional knowledge on $E$ can 
be expected from exclusive vector meson production.

Running with a polarized internal target is presently still 
technologically disfavored, because the maximum achievable 
instantaneous luminosity falls short by about two orders of magnitude 
compared to unpolarized targets. In the case of HERMES running 
beyond 2006, a polarized target with a density improvement by about a 
factor of two may become available. 
\begin{figure}[htb]
\centering
\vspace{-1.2cm}
\includegraphics[width=7.5cm]{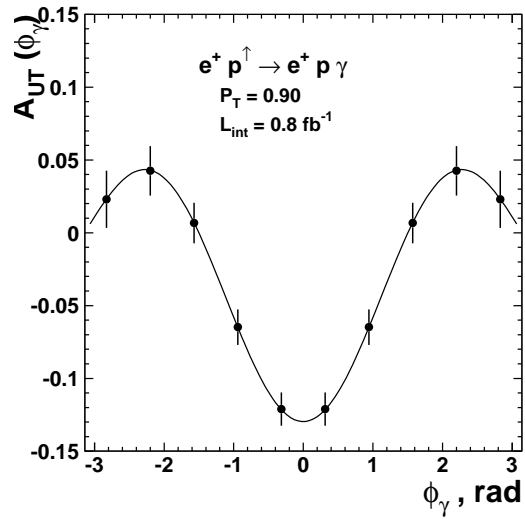}
\vspace{-1.5cm}
\caption{\it Projected statistical accuracy for the azimuthal
             dependence of the DVCS asymmetry in a low-luminosity
             experiment with unpolarized electron beam and 
             transversely polarized target \cite{VKpriv}, based 
             upon the HERMES acceptance.}
\label{fig:TpolDVCSasy}
\vspace{-0.7cm}
\end{figure}

From Fig.~\ref{fig:TpolDVCSasy} it can be seen that the statistical
precision, achievable after three years of HERMES running with an 
unpolarized beam and a transversely polarized target,
would be sufficient to measure the azimuthal dependence of the DVCS
asymmetry $A_{UT}$ \cite{VKpriv}. The curve
depicts the behaviour of the chosen GPD parameterization (B), as 
explained in detail in Ref.~\cite{KoNoDVCS}. Note that its missing 
periodicity in $\phi_\gamma$ is due to a particular way of integrating 
over one of the azimuthal angles in Ref.~\cite{BeMueNieSch}.
It remains to be shown that, even using the approximation of $u$-quark 
dominance, this data set would allow simultaneous access to the
GPDs $H$ and $E$. Extrapolating them to $t=0$ 
and simultaneously calculating the integral over $x$ would make it possible
in principle to obtain a first
glimpse of the total angular momentum carried by $u$-quarks, $J_u$.
However, this may not be feasible because of lack of statistics and
several theoretical uncertainties that have to be quantified reliably. \\

A real breakthrough towards a determination of GPDs can be expected 
only from high-statistics measurements at a future high-luminosity 
facility, as outlined above. A typical rather conservative assumption 
on integrated luminosity is 100 fb$^{-1}$ per year of running (cf.
Ref.~\cite{TESLA-N}). As explained above, a key experiment would 
consist in measuring DVCS with an unpolarized beam and a transversely 
polarized target. In this case a typical observable would be the 
$\cos{(\phi /2)}$ moment of the cross section asymmetry that
projects out the set of GPDs as discussed above. 

\begin{figure}[ht]
\centering
\vspace{-0.5cm}
\includegraphics[width=7.5cm]{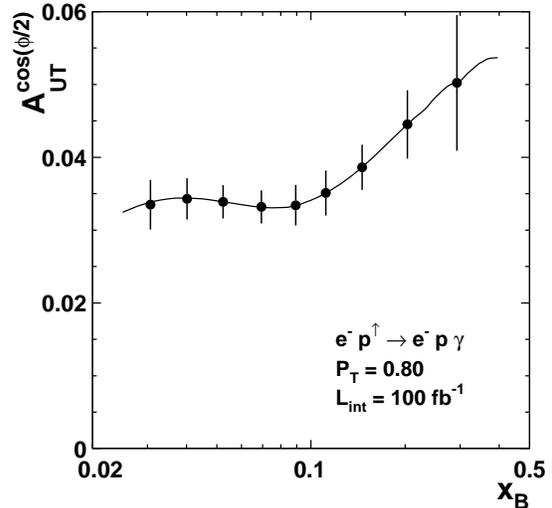}
\vspace{-1.6cm}
\caption{\it Projected statistical accuracy for the $x_B$-dependence
             of the $\cos{(\phi /2)}$ moment in a high-luminosity
             experiment with unpolarized electron beam and transversely
             polarized target \cite{VKpriv}, using
             a conical acceptance between 40 and 220 mrad.}
\label{fig:HighLumiDVCSasy}
\vspace{-1cm}
\end{figure}

In Fig.~\ref{fig:HighLumiDVCSasy} the statistical accuracy is illustrated 
for a measurement of DVCS in a 30 GeV electron beam with a high-resolution 
spectrometer, combined with a NH$_3$-target (80\% polarization, 
dilution factor 17.6\%) and applying the same 
cuts as used in the analysis of Ref.~\cite{HermesDVCS}. The error bars 
show the projected statistical 
accuracy achievable in 1 year of running for the $x_B$-dependence of 
the moment $A^{\cos{(\phi /2)}}_{UT}$ introduced above. The curve
again represents the behaviour of the GPD parameterization (B) 
\cite{KoNoDVCS}. The statistical errors
clearly show that an acceptable precision can be reached and that even
the measurement of 2-dimensional dependences (e.g. on $x_B$ and $t$)
are within reach. It remains to be shown by a careful analysis of 
simulated data that the ultimate goal, the above envisioned `complete' 
determination of the angular momentum structure of the nucleon, can be 
experimentally achieved
and that the most complex variety of theoretical uncertainties can be 
brought under control to make this high-statistics result meaningful.

\section{SUMMARY AND CONCLUSION}

In the context of the proposed TESLA/FEL project, a new
high-luminosity high-resolution electron-nucleon facility may be 
accommodated. This would pave the way for a great variety of measurements, 
ranging from inclusive scattering through semi-inclusive final states to 
hard exclusive processes. On the basis of the anticipated statistical 
accuracies, shown for various key experiments, it appears justified to 
expect that many questions on the structure of hadronic matter, 
especially on its momentum and spin structure, can be successfully answered. 

It goes without saying that there will be lots of further extremely 
interesting physics results beyond those discussed above.
Among the published HERMES results, after 5 years of running, only 
10-20\% had been anticipated in the proposal.


\section*{ACKNOWLEGDEMENTS}

I am deeply indebted to V.A.~Korotkov for calculating new DVCS 
projections, to A.~Miller for a very careful reading of the manuscript
and to D.~von~Harrach for an update of Fig.~\ref{fig:resolELFE}.
Many thanks to them and to S.~Bass, M.~Diehl, P.~Hoyer, D.~M\"uller, 
D.~Ryckbosch and G.~van~der~Steenhoven for very valuable comments. The
help of R.~Kaiser in the preparation of Fig.~\ref{fig:lumi_E} is 
acknowledged.

\end{document}